\documentclass[twocolumn,pra,showpacs,superscriptaddress]{revtex4}
\usepackage{ulem}
\usepackage{amssymb}
\usepackage{amsmath}
\usepackage{graphicx}
\usepackage{subfigure}
\usepackage{natbib}
\usepackage{epsfig}
\usepackage{amsfonts}
\usepackage{mathrsfs}
\usepackage{CJK}
\usepackage{subfigure}
\usepackage[toc,page,title,titletoc,header]{appendix}
\usepackage{color}

\begin{document}

%\begin{CJK*}{GBK}{song}

\title{Stable Heteronuclear Few-Atom Bound States in Mixed Dimensions}

\author{Tao Yin}
\author{Peng Zhang}
\email{pengzhang@ruc.edu.cn}%
\author{Wei Zhang}
\email{wzhangl@ruc.edu.cn}%
\affiliation{Department of Physics, Renmin University of China,
Beijing 100872, People's Republic of China}%

\begin{abstract}
We study few-body problems in mixed dimensions with $N \ge 2$ heavy atoms trapped
individually in parallel one-dimensional tubes or two-dimensional disks, and a single light
atom travels freely in three dimensions. By using the Born-Oppenheimer approximation,
we find three- and four-body bound states for a broad region of heavy-light atom scattering length
combinations. Specifically, the existence of trimer and tetramer states persist to negative
scattering lengths regime, where no two-body bound state is present. These few-body bound
states are analogous to the Efimov states in three dimensions, but are stable against three-body
recombination due to geometric separation. In addition, we find that the binding energy of the
ground trimer and tetramer state reaches its maximum value when the scattering lengths are
comparable to the separation between the low-dimensional traps. This resonant behavior is
a unique feature for the few-body bound states in mixed dimensions.
\end{abstract}

\maketitle
%\end{CJK*}

\section{Introduction}

One striking feature of few-body physics is the presence of universality
under a resonant short-range interaction,
where the low-energy behavior of the system does not depend
on the details of its structure or interactions at short distances.
Of particular interest is the existence of bound trimer states for three identical bosons
in three dimensions with a resonant two-body interaction, as discussed in 1970 by
Vitaly Efimov~\cite{efimov-70}. At infinite scattering length, these three-body bound states form
an infinite geometric spectrum with a constant ratio between two successive binding energies, indicating a
discrete scaling symmetry~\cite{tolle-11}. Besides, the bound trimer states
persist rather counterintuitively to negative scattering length regime,
where two-body bound states are not existent.
After its original proposal, the Efimov physics has attracted great attention in multi-disciplinary
systems, including atomic nuclei~\cite{jensen-04, mazumdar-06},
$^4$He trimers~\cite{lim-77, bruhl-05}, and other molecules~\cite{baccarelli-00}.
However, a direct evidence of such peculiar behavior was not achieved for more
than three decades until its first observation in an ultracold gas of neutral atoms~\cite{kraemer-06}.
Thanks to the extraordinary controllability of the mutual atomic interaction by
tuning through a magnetic Feshbach resonance, signatures of trimer bound states
have been observed in trapped atomic gases for both negative and positive
scattering length regimes~\cite{kraemer-06, knoop-09, zaccanti-09, gross-09, pollack-09}.

In addition to the original problem of identical bosons, the study of three-body
physics has been extended to a variety of other three-particle
systems~\cite{braaten-06}, including three distinguishable
particles with different scattering lengths and/or different masses~\cite{amado-72, efimov-72,Fonseca},
two identical fermions with a third atom~\cite{efimov-72, petrov-03, petrov-04},
and the three-atom
systems with with non-zero angular momentum~\cite{endo-11}.
Of particular interest is the case of three distinguishable fermions in an ultracold
gas of three-component $^6$Li atoms. In such a system, there exits a broad
magnetic Feshbach resonance such that all three scattering lengths can be tuned
around resonance simultaneously~\cite{ottenstein-08},
leading to a promising candidate to observe the few-body universal
behavior~\cite{huckans-09, williams-09, braaten-09, wenz-09, nakajima-10, naidon-11, nakajima-11, lompe-10a, lompe-10b}.
Besides, the few-body problem has also been analyzed in ultracold gases of
different atomic species by tuning the interaction across an interspecies Feshbach
resonance~\cite{barontini-09}.

Due to the multi-channel nature of the inter-atomic interaction, the Efimov states
in the three-dimensional (3D) ultracold atomic gases are only metal-stable states. Through the
three-body recombination process, two of
the three atoms
in an Efimov trimer can further form a deeply bound dimer and the third one would escape from the trap.
In order to prepare stable trimer states, one has to figure out a mechanism to
significantly reduce or even prevent three-body recombination.
Since the three-body recombination process only occurs when three atoms all come to
a close range, one possible route towards this goal is to use geometric confinement
to separate atoms such that they can not travel to a same spot.
For instance, if two of the three atoms are individually trapped in two spatially separated
one-dimensional (1D) tubes or two-dimensional (2D) disks, and interact with each other
via the third atom which is free in all three dimensions (3D), the three-body recombination
is inherently forbidden and the trimer states are stable
if they exist in this mixed dimensional configuration.

The few-body problem in mixed dimensions has been recently discussed by
Nishida and Tan~\cite{nishida-08}, where they consider two species of atoms
confined in different dimensions and find trimer bound states for a certain range of
mass ratio. However, since the atoms in lower dimensions are not geometrically
separated, this configuration suffers the same problem of three-body recombination
and the trimer states are unstable. Therefore, Nishida consider the problem of
two atoms trapped in two separated 1D tubes or 2D layers, and interacts with
the third atom which is free in 3D~\cite{nishida-10, nishida-11}.
This 1D-1D-3D or 2D-2D-3D mixture thus can support
stable Efimov trimer states.

In this manuscript, we adopt another approach based on Born-Oppenheimer
approximation (BOA) to study some few-body problems in
mixed dimensions, and investigate the existence and properties of stable few-body
bound states in a variety of configurations. For the three-body problems, we consider
the systems with two heavy atoms trapped in two parallel 1D tubes
(1D-1D-3D) or 2D disks (2D-2D-3D),
plus one light  atom moving freely in 3D (see Fig. 1 for illustration).
We conclude that the light atom can induce an effective interaction between the two heavy
atoms which are spatially separated by the low dimensional traps.
Due to this effective interaction, the two heavy atoms can be bound with each other
and lead to the formation of a three-body bound state in a very broad parameter region,
including the regimes with negative
$s$-wave scattering lengths between the light and the two heavy atoms, where
two-body bound states are not present. 
%With an analytical expression for the effective interaction, we can
%obtain the bound states of the three-body problem numerically.
%For both 1D-1D-3D and 2D-2D-3D geometries, we find universal three-body bound states
%living in a wide range of scattering length, in particular within the regimes with negative
%$s$-wave scattering lengths between the light and the two heavy atoms, where
%two-body bound states are not present.

In addition to their existence in mixed dimensions, the universal three-body
bound states also acquire some unique features due to the geometric confinement.
Especially, we find that the two heavy atoms experience the strongest effective interaction
when the scattering length between heavy and light atoms equals to the distance
between the two low-dimensional traps. As a consequence of this
resonance phenomenon, the binding energy of the ground trimer state
takes a peak value around the resonance point, where the scattering
length is of a finite value. We emphasis that, the BOA provides a
very clear physical picture with which the new resonance phenomenon in the
mixed-dimensional systems can be easily explored and clearly described.

We also compare our results with the
exact expression~\cite{nishida-10, nishida-11} given by an effective field theory, and conclude
that BOA works well even in systems with a mass ratio only about $6$. This finding
suggests that BOA is a powerful tool for the study of stable heteronuclear
few-body bound states in mixed dimensions.

To demonstrate the usage of BOA for general few-body problems,
we consider as an example the 1D-1D-1D-3D geometry with three heavy atoms
confined individually in parallel 1D tubes and a light atom in 3D free space. We find
four-body bound states living in a wide range of scattering lengths.
A similar resonance phenomenon is also observed when the scattering length
becomes close to the mutual distances between 1D tubes, in which case the
binding energy of the ground tetramer state reaches its maximum when the three
1D tubes form an equilateral triangle. We also show the scheme to generalize
the BOA to problems with $N \ge 3$ heavy atoms and a single light one in an
arbitrary mixed dimensional geometry. The mixed dimensional systems discussed
in this manuscript can be realized in a mixture of two-species ultracold gases with
species-selective dipole potential, as illustrated in recent experiments~\cite{catani-09}.

The remainder of this manuscript is organized as follows. In Sec. II we first consider
the 1D-1D-3D geometry and outline the BOA approach for the three-body problem.
We calculate the effective interaction potential between the two heavy atoms,
and observe the new resonance phenomenon. In Sec. III we solve for
the three-body bound states, from which we conclude that
a stable trimer state can exist in a broad parameter region, and the binding energy
of the ground trimer state takes largest value under the new resonance condition.
Similar results of the 2D-2D-3D system are shown in Sec. IV.
In Sec. V we extend the usage of BOA to the four-body problem in
1D-1D-1D-3D geometry, and discuss the existence and properties of bound
tetramer states. In Sec. VI, we show the general scheme to apply BOA
in problems with more than 3 atoms in arbitrary mixed dimensional geometries.
Our main findings are concluded in Sec. VII, and the Bethe-Peierls boundary
condition used in our BOA approach is derived in Appendix A.

%%%%%%
\section{BOA for three-atom bound states in 1D-1D-3D systems}

In this section we present the Born-Oppenheimer approach for a
three-body system with two heavy atoms individually trapped in
two parallel 1D tubes and a light atom moving freely in the 3D space.
The straightforward generalization to 2D-2D-3D systems will be given
in Sec. IV, while the discussion for four-body problems in 1D-1D-1D-3D systems
is given in Sec. V.

\subsection{System and Hamiltonian}

As shown in Fig. 1(a), the 1D-1D-3D system includes two heavy atoms $A_{1}$ and
$A_{2}$, plus a light atom $B$. The atoms $A_{1}$ and $A_{2}$ are trapped in two
parallel 1D tubes centered along the lines $\left( x= \pm L/2,y=0\right)$,
while the light atom $B$ moves freely in the 3D space.
The quantum state of this system can be described by the wave
function $\Psi (\vec{r}_{B};z_{1},z_{2})$, where $z_{1,2}$ are the $z$-coordinate
of atoms $A_{1,2}$ in the 1D tubes, and $\vec{r}_{B}=(x_{B},y_{B},z_{B})$
is the coordinate of atom $B$ in 3D.

\begin{figure}[tbp]
\includegraphics[bb=52bp 525bp 513bp 755bp,clip,width=8cm]{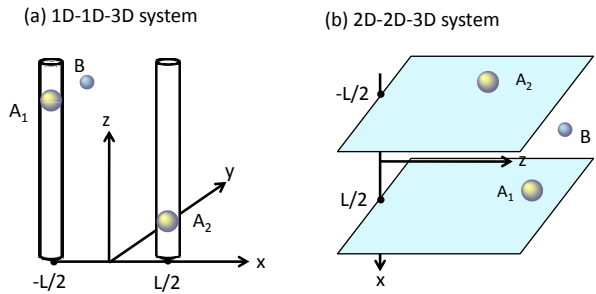}
\caption{(color online) (a) The 1D-1D-3D system with two heavy atoms $A_1$ and $A_2$
confined in two 1D tubes and the light atom $B$ moving freely in 3D.
(b) The 2D-2D-3D system with two heavy atoms $A_1$ and $A_2$
confined in two 2D planes and the light atom $B$ moving freely in the 3D space.}
\end{figure}

In this manuscript, we use the natural units $\hbar =m_{B}=L=1$, where $m_{B}$ is
the mass of atom $B$. The Hamiltonian for the motion of the three atoms is
\begin{eqnarray}
H=-\frac{1}{2}\nabla _{B}^{2}-\frac{1}{2m_{1}}\frac{\partial ^{2}}{\partial
z_{1}^{2}}-\frac{1}{2m_{2}}\frac{\partial ^{2}}{\partial z_{2}^{2}}+V_{1B}+V_{2B},
\label{hf}
\end{eqnarray}
where $m_{1,2}$ are the masses of atoms $A_{1,2}$ in the natural unit,
and $V_{1B,2B}$ are the interaction potentials between $A_{1,2}$ and $B$, respectively.
In this work we only consider the cases where the distance $L$ between the two tubes
is much larger than the characteristic length of the interaction potential between $A_{1}$
and $A_{2}$. Hence the $A_{1}$-$A_{2}$ interaction can be safely ignored.

\subsection{BOA for three-body bound states}

The three-body bound state is given by the solution of the eigen-equation
\begin{eqnarray}
H\Psi =E\Psi.  \label{ee}
\end{eqnarray}
When the masses of the heavy atoms $A_{1,2}$ is much larger
than the one of $B$, or $m_{1,2} \gg 1$ in the natural unit,
the eigen-equation (\ref{ee}) can be solved with
BOA. This approximation is applicable when the motion
of the heavy atoms $A_{1,2}$ is slow enough such that the quantum transitions
between different instantaneous eigen-states of the light atom $B$ with
fixed positions $z_{1,2}$ of $A_{1,2}$ are negligible. Therefore, the total wave function $\Psi $
of the three-body bound state can be approximated as a factorized form
\begin{eqnarray}
\Psi (\vec{r}_{B};z_{1},z_{2})=\phi (z_{1},z_{2})
\psi (\vec{r}_{B},z_{1},z_{2}), \label{pf2}
\end{eqnarray}
where $\psi (\vec{r}_{B},z_{1},z_{2})$ is an instantaneous bound-state
solution of the eigen-equation of the Hamiltonian of atom $B$ with fixed values of $z_1$
and $z_2$.

As shown in Appendix A, we can further replace the interaction potentials $V_{1B}$
and $V_{2B}$ with the Bethe-Peierls boundary conditions
\begin{eqnarray}
\psi (r_{1B} \to 0)\propto \left( 1-\frac{a_{1}}{r_{1B}}\right) +%
\mathcal{O}(r_{1B});  \label{c1} \\
\psi (r_{2B} \to 0)\propto \left( 1-\frac{a_{2}}{r_{2B}}\right) +%
\mathcal{O}(r_{2B}).  \label{c2}
\end{eqnarray}%
Here $r_{1B,2B}$ are the relative distances between the heavy atoms $A_{1,2}$ and the
light atom $B$, $a_{1,2}$ are the mixed-dimensional scattering lengths between
$A_{1,2}$ and $B$. Notice that the Bethe-Peierls boundary conditions
(\ref{c1}) and (\ref{c2}) are derived from the first-principle calculation
where the $3$D motion of all the three atoms $A_{1,2}$ and $B$ are taken into account.
Then the mixed-dimensional scattering lengths $a_{1,2}$ are
determined by both the 3D $s$-wave scattering lengths between $A_{1,2}$ and $B$,
as well as the intensity of the transverse confinements of the 1D traps.
Thus, $a_{1,2}$ can be tuned either through a 3D magnetic Feshbach resonance~\cite{Feshbach}
or via a mixed-dimensional confinement-induced resonance~\cite{nishida-08}.

With the Bethe-Peierls boundary conditions, the
wave function $\psi (\vec{r}_{B},z_{1},z_{2})$ is determined by
\begin{eqnarray}
-\frac{1}{2}\nabla _{B}^{2}\psi (\vec{r}_{B},z_{1},z_{2})
=
V_{\rm{eff}}(z_{1},z_{2})\psi (\vec{r}_{B},z_{1},z_{2}),  \label{ee3}
\end{eqnarray}
through which the shape of the wave function $\psi (\vec{r}_{B},z_{1},z_{2})$
and the relevant eigen-energy $V_{\rm{eff}}(z_{1},z_{2})$ can be determined
for a given value of $z_{1,2}$.

In the approach of BOA, the instantaneous energy $V_{\rm{eff}}(z_{1},z_{2})$
of the light atom $B$ serves as an effective potential between the two slowly
moving heavy atoms. Then the wave function $\phi(z_{1},z_{2}) $ in Eq. (\ref{pf2})
satisfies the Schr{\"o}dinger equation
\begin{eqnarray}
\left[ -\frac{1}{2m_{1}}\frac{\partial ^{2}}{\partial z_{1}^{2}}
-\frac{1}{2m_{2}}\frac{\partial ^{2}}{\partial z_{2}^{2}}
+V_{\rm{eff}}(z_{1},z_{2})\right] \phi (z_{1},z_{2})\nonumber\\
=E\phi (z_{1},z_{2}),
\label{eef}
\end{eqnarray}
where $E$ is the total energy of the trimer state defined in Eq. (\ref{ee}).
In this manuscript, we focus only on the ground state of the three-body
eigen-equation (\ref{ee}), which is consisted of the ground-state
solutions $\psi $ and $\phi$ of (\ref{ee3}) and (\ref{eef}), respectively.

In summary, to derive the three-body bound state with BOA, we should first
find the ground-state solution $\psi $ of the instantaneous eigen-equation (\ref{ee3})
of the light atom $B$, and then solve the effective
eigen-equation (\ref{eef}) of the heavy atoms $A_{1,2}$ where the
instantaneous eigen-energy $V_{\rm{eff}}(z_{1},z_{2})$ of $\psi$ plays a role
as interaction potential between $A_{1}$ and $A_2$. Therefore, the BOA provides
a simple and clear physical picture for the three-body problem, i.e., the light atom $B$
induces an effective interaction between the two heavy atoms,
which determines the properties of the three-body bound state.
With this picture, one can perform
not only quantitative calculations but also qualitative discussions
for the appearance and features of the trimer states when the
potential function $V_{\rm{eff}}(z_{1},z_{2})$ is known from (\ref{ee3}).
This is a major advantage of the BOA approach.

In the end of this subsection we emphasize that, since in BOA the
transitions between different solutions of the instantaneous
eigen-equation (\ref{ee3}) is neglected, this approximation
can only be used when the gap between $V_{\rm{eff}}(z_{1},z_{2})$ and other
eigen-energies of (\ref{ee3}) [with boundary conditions (\ref{c1}) and
(\ref{c2})] is large enough. In the cases where $V_{\rm{eff}}(z_{1},z_{2})$
is close to the lower bound of the continuous spectrum, the application
of BOA may be questionable.

%%%%%%%%%%%%%%
\subsection{Effective interaction between the two heavy atoms}

In the discussion above we outline the procedure for the derivation of the
three-body bound states with BOA. In this subsection we solve
Eqs. (\ref{c1}-\ref{ee3}) to calculate the instantaneous eigen-state
$\psi (\vec{r}_{B},z_{1},z_{2})$ of the light atom $B$, and the light-atom-induced
effective potential $V_{\rm{eff}}(z_{1},z_{2})$ between the two heavy atoms.

A straightforward calculation shows that the lowest ground state solution $\psi$
(up to a normalization factor) of Eq. (\ref{ee3})
and the corresponding energy $V_{\rm{eff}}(z_{1},z_{2})$ are
given by
\begin{eqnarray}
\psi (\vec{r}_{B},z_{1},z_{2}) &=&\frac{e^{-\kappa (r_{12})r_{1B}}}{r_{1B}}
+\xi (r_{12})\frac{e^{-\kappa (r_{12})r_{2B}}}{r_{2B}}  \label{psi1}; \\
V_{\rm{eff}}(z_{1},z_{2}) &=&-\frac{\kappa ^{2}(r_{12})}{2},\label{v1}
\end{eqnarray}
where $r_{12}=\sqrt{1+(z_{1}-z_{2})^{2}}$ is the distance
between $A_{1}$ and $A_{2}$.

Substituting the expression of $\psi (\vec{r}_{B},z_{1},z_{2})$ into the
Bethe-Peierls boundary conditions (\ref{c1}) and (\ref{c2}), one can derive
the values of $\kappa $ and $\xi $ in terms of the distance $(z_{1}-z_{2})$,
and then obtain expressions for $\psi (\vec{r}_{B},z_{1},z_{2})$
and $V_{\rm{eff}}(z_{1},z_{2})$.
Notice that, as a bound state, the wave function
$\psi (\vec{r}_{B},z_{1},z_{2})$ must approach zero in the limit
$r_{1B}\to \infty $ or $r_{2B}\to \infty $. Therefore, the
condition $\kappa >0$ must be satisfied when we solve the equations
of $\kappa $ and $\xi $.

According to Eq. (\ref{v1}), the effective potential $V_{\rm{eff}}(z_{1},z_{2})$
is a function of distance $z_{12} \equiv z_{1}-z_{2}$ between the
two heavy atoms along the axial direction of 1D tubes. Then the wave
function $\phi (z_{1},z_{2})$ in the total wave function (\ref{pf2}) is also
a function of $z_{12}$, indicating the translational symmetry along the $z$-axis.
From now on, we rewrite $V_{\rm{eff}}(z_{1},z_{2})$
as $V_{\rm{eff}}(z_{12})$, and $\phi (z_{1},z_{2})$ as $\phi (z_{12})$,
and write Eq. (\ref{eef}) as
\begin{eqnarray}
\left[ -\frac{1}{2m_{\ast }}\frac{\partial ^{2}}{\partial z_{12}^{2}}
+V_{\rm{eff}}(z_{12})\right] \phi \left( z_{12} \right) =E\phi \left(z_{12}\right),  \label{ef2}
\end{eqnarray}
where $m_{\ast }=m_{1}m_{2}/(m_{1}+m_{2})$
is the reduced mass of the two heavy atoms.
From Eq. (\ref{ef2}), we can see clearly that $V_{\rm{eff}}$
serves as an effective interaction between the two heavy atoms $A_{1,2}$,
and determines the existence and behavior of the three-body bound states.
Next, we discuss the feature of $V_{\rm{eff}}$ in different parameter regions.

\subsubsection{$a_{1}=a_{2}=a>0$}

In this case the two heavy atoms $A_{1,2}$ have the same positive scattering
length with the light atom $B$. Since $\psi $ is the ground-state solution
of Eq. (\ref{ee3}), a straightforward calculation shows that in this
symmetric case we have $\xi =1$ and $\kappa $ given by the equation
\begin{eqnarray}
-\kappa +\frac{e^{-\kappa r_{12}}}{r_{12}}=-\frac{1}{a}.  \label{ke}
\end{eqnarray}%
This equation can be solved analytically, leading to,
\begin{eqnarray}
\kappa =\frac{1}{a}+\frac{W\left( e^{-r_{12}/a}\right) }{r_{12}},  \label{k1}
\end{eqnarray}%
where $W(z)$ is Lambert $W$ function or the principle root of
equation $z=We^{W}$. Substituting the result (\ref{k1}) into Eq. (\ref{v1}),
we finally obtain an analytic expression of the effective interaction
between the two heavy atoms:%
\begin{eqnarray}
V_{\rm{eff}}(z_{12})=U\left( a;z_{12}\right) -\frac{1}{2a^{2}},  \label{vv1}
\end{eqnarray}%
where the regularized part $U\left( a;z_{12}\right) $ is given by
\begin{eqnarray}
U\left( a;z_{12}\right) &=&-\frac{1}{2}
\frac{W\left( e^{-\sqrt{1+z_{12}^{2}}/a}\right)^{2}}
{1+z_{12}^{2}}
\nonumber \\
&&\hspace{-0.5cm}
-\frac{1}{a}\frac{W\left( e^{-\sqrt{1+z_{12}^{2}}/a}\right) }
{\sqrt{1+z_{12}^{2}}},
\label{u1}
\end{eqnarray}
which approaches zero in the limit $\left\vert z_{12}\right\vert \rightarrow \infty $.
Therefore, the characters of bound states are essentially determined by the
behavior of $U\left( a;z_{12}\right)$.

\begin{figure}[tbp]
\includegraphics[bb=33bp 277bp 464bp 572bp,clip,width=8cm]{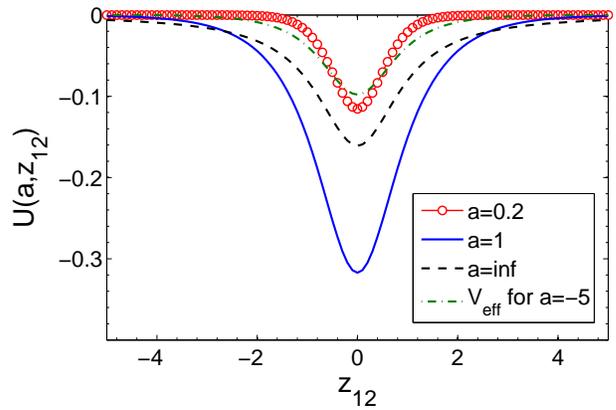}
\caption{(color online) The regularized effective potential $U(a,z_{12})$
between the two heavy atoms $A_{1,2}$ with scattering lengths
$a_1=a_2=a=0.2$ (red solid line with open circles), $1$ (blue solid line)
and $\infty$ (black dashed line). We also plot the effective potential
$V_{\rm eff}(z_{12})$ for $a_1=a_2=-5$ (green dashed-dotted line).
The natural unit of $\hbar = m_B = L =1$ is used throughout this paper.
}
\label{uz1z2}
\end{figure}

With the knowledge of the $W$ function, we can easily find that when $a>0$,
$U\left( a;z_{12}\right)$ is a pure symmetric potential well with
\begin{eqnarray}
U\left( a,z_{12}\right) =U\left( a,-z_{12}\right) <0.
\end{eqnarray}
In Fig.~\ref{uz1z2}, we plot $U\left( a;z_{12}\right) $ for a set of typical
values of scattering lengths. It is clearly shown that
$U\left( a;z_{12}\right) $ provides a simple $1$D potential well for the two heavy atoms.
This behavior guarantees that there exists at least one bound-state
solution $\phi$ of Eq. (\ref{ef2}), and then the total system has at least one
three-body bound state.

Intuitively speaking, one would expect that the atom-atom interaction
effect be most significant when the scattering length takes infinite value.
However, we find from Fig. 2 that the depth of the effective interaction
$U\left( a;z_{12}\right) $ takes a maximum value when $a=1$ in our natural unit,
rather than $a \to +\infty $. This observation suggests that the light-atom-induced
interaction between the two heavy atoms $A_{1,2}$ is most significant when
the scattering length between a single heavy atom and the light one
equals to the distance separating the two 1$D$ tubes. This novel property
can be considered as a kind of resonance effect given by the special
configuration of mixed dimensional systems.

\begin{figure}[tbp]
\includegraphics[bb=34bp 271bp 446bp 572bp,clip,width=8cm]{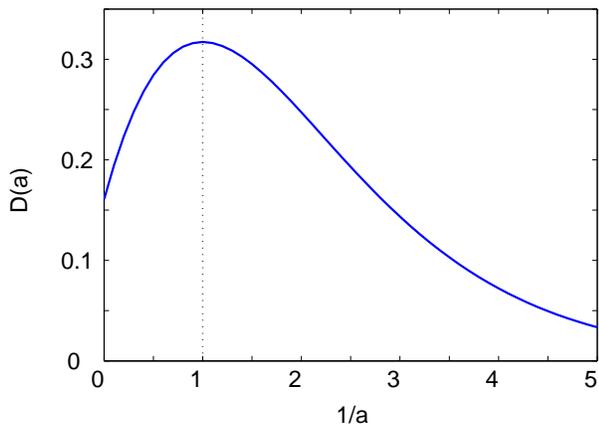}
\caption{(color online) The depth $D(a)$ of the regularized part $U(a,z_{12})$
of the effective interaction between the two heavy atoms in the case of $a_1=a_2=a>0$.
Notice that $D(a)$ takes a maximum value at $a=1$, indicating a new
resonance behavior for mixed dimensional systems.}
\label{depth}
\end{figure}

This resonance effect can also be proved analytically with the character
of the $W$ function. For any given value of $a$, the potential
$U\left(a,z_{12}\right) $ has only one minimum point,
which is localized at the origin $z_{12}=0$.
Thus, the depth of the potential well takes the form
\begin{eqnarray}
D\left( a\right)  \equiv -U\left( a,0\right) =\frac{1}{2}W\left( e^{-1/a}\right)
^{2}+\frac{1}{a}W\left( e^{-1/a}\right).
\end{eqnarray}%
It is easy to show that $D\left( a\right) $ takes the maximum value when
$a=1$. In Fig. \ref{depth} we plot the potential depth as a function of
$1/a$, exhibiting the resonance signature at $a = 1$.

\subsubsection{$a_{1}=a_{2}=a<0$}

In this case, by substituting Eq. (\ref{psi1}) into the Bethe-Peierls
boundary conditions (\ref{c1}) and (\ref{c2}), we also get $\xi =1$ and
$\kappa $ given by Eqs. (\ref{ke}) and (\ref{k1}) for $r_{12}<\left\vert a\right\vert $.
However, for $r_{12}>\left\vert a \right\vert $, there is no positive solution
of Eq. (\ref{ke}) for $\kappa $. This suggests that the Schr\"{o}dinger equation (\ref{ee3})
with Bethe-Peierls boundary conditions (\ref{c1}) and (\ref{c2}) do not support any
instantaneous bound state $\psi $ of the light atom $B$, and then one
cannot derive any effective interaction for the two heavy atoms $A_{1,2}$ within BOA.
As a consequence, when the scattering length $|a| < 1$, there would be no three-body bound
state since the condition $r_{12} > |a|$ is satisfied with arbitrary 1D distance $z_{12}$
between the two heavy atoms.

On the other hand, when $\left\vert a \right\vert >1$,
the BOA can give the effective interaction potential
\begin{eqnarray}
V_{\rm{eff}}\left(z_{12}\right) =-\frac{1}{2}\left[ \frac{1}{a}
+\frac{W\left(e^{-\sqrt{1+z_{12}^{2}}/a}\right) }{\sqrt{1+z_{12}^{2}}}\right] ^{2},
\end{eqnarray}
provided that $|z_{12}| < \sqrt{a^{2}-1}$ or $r_{12}<\left\vert a\right\vert $.
In the outer region of $|z_{12}| > \sqrt{a^{2}-1}$, the potential takes zero value
as $V_{\rm{eff}}\left( z_{12}\right) =0$. In Fig. \ref{uz1z2},
we also show $V_{\rm{eff}}\left( z_{12}\right) $ with negative scattering length.

We would like to emphasize that BOA can only be used when $V_{\rm{eff}}$ is
well-separated from the continuous spectrum of the Schr\"{o}dinger equation (%
\ref{ee3}). This criteria is actually broken in the region
$r_{12} \sim \left\vert a\right\vert $ or $z_{12} \sim \sqrt{a^{2}-1}$,
where we have $V_{\rm{eff}}\left( r_{12}\right) \sim 0$.
Then the effective potential is not applicable in these regions.
Fortunately, if the potential is deep enough, the ground-state
wave function $\phi $ of the heavy atoms $A_{1,2}$ would be
mainly localized in the region $z_{12} \sim 0$ or $r_{12} \ll \left\vert a\right\vert $,
where BOA is applicable. Thus, the ground-state wave function
and its corresponding binding energy obtained from BOA is still
reliable. Notice that in this negative scattering length regime,
$A_{1,2}$ and $B$ cannot form any two-body bound state,
hence the appearance of a three-body bound state is a non-trivial
universal phenomenon.

\subsubsection{$0<a_{1}<a_{2}$ or $a_{2}<0<a_{1}$}

Now we consider the general cases where the scattering lengths $a_1$ and $a_2$
are different. In these cases one can also derive the values of $\xi $ and $\kappa $ by substituting
the expression (\ref{psi1}) into the Bethe-Peierls boundary conditions (\ref%
{c1}) and (\ref{c2}). When $0<a_{1}<a_{2}$ or $a_{2}<0<a_{1}$, we know that
in the limit $r_{12}\rightarrow \infty $, that is the two heavy atoms are far away
from each other, the instantaneous ground state of the light atom $B$ is the
two-body bound state of $B$ and $A_{1}$. Considering the expression
(\ref{psi1}) of the instantaneous bound state, we have%
\begin{eqnarray}
\xi \left( r_{12}\rightarrow \infty \right) =0.
\end{eqnarray}%
With the help of this condition, we obtain the result%
\begin{eqnarray}
\xi =\frac{-\Delta +\sqrt{\Delta ^{2}+4e^{-2\kappa r_{12}}/r_{12}^{2}}}{2}%
r_{12}e^{\kappa r_{12}},  \label{xie}
\end{eqnarray}%
where%
\begin{eqnarray}
\Delta \equiv \frac{1}{a_{1}}-\frac{1}{a_{2}}>0.\label{delta}
\end{eqnarray}%
Then the value of $\kappa $ is given by
\begin{eqnarray}
-\kappa +\frac{-\Delta +\sqrt{\Delta ^{2}+4e^{-2\kappa r_{12}}/r_{12}^{2}}}{2%
}=-\frac{1}{a_{1}}.  \label{kkkq}
\end{eqnarray}%
By solving Eqs. (\ref{xie}) and (\ref{kkkq}) numerically, we can obtain the
values of $\xi $ and $\kappa $, and then the effective potential
$V_{\mathrm{eff}}$. It is easy to show that $V_{\rm eff}<0$ for
all values of $z_{12}$. Therefore, there is also at least one
three-body bound state. When $a_{2}<0<a_{1}$,
although the atoms $A_{1}$ and $B$ can form a two-body bound state,
there is no two-body bound state for $A_{2}$ and $B$. In this sense
the existence of a three-body bound state is also a non-trivial phenomenon.

\subsubsection{$a_{1}<a_{2}<0$}

In this case a straightforward calculation shows that the values of $\xi $
and $\kappa $ are also determined by Eqs. (\ref{xie}) and (\ref{kkkq}).
Nevertheless, similar to the case of $a_{1}=a_{2}=a<0$,
there are also some regions where the instantaneous bound state
$\psi $ does not exist. Specifically, we can define a
critical distance $r_{12}^{\ast }$ as
\begin{eqnarray}
r_{12}^{\ast }=2\left[ \left( \Delta -\frac{2}{a_{1}}\right) ^{2}-\Delta^{2}\right] ^{-1/2}
\end{eqnarray}%
with $\Delta$ defined in (\ref{delta}).
It is apparent that when $r_{12}>r_{12}^{\ast }$, we cannot find any real
$\kappa $ which satisfies Eq. (\ref{kkkq}).
In this sense, $r_{12}^{\ast}$ can be understood as the range of the
effective interaction between $A_1$ and $A_2$.
When this range is smaller than the distance between the two 1D tubes,
i.e. $r_{12}^{\ast} < 1$, the two heavy atoms are always separated
far enough such that the BOA does not give any effective mutual interaction.
On the other hand, when $r_{12}^{\ast }>1$ the effective potential
of the two heavy atoms can be defined as%
\begin{eqnarray}
V_{\rm eff}=\left\{
\begin{array}{c}
-\kappa^{2}/2;\ \ \ 1\leq r_{12}\leq r_{12}^{\ast } \\
0;\ \ \ r_{12}>r_{12}^{\ast }.%
\end{array}%
\right.
\end{eqnarray}%
This potential is also not reliable in the region $r_{12}\sim r_{12}^{\ast }$
where the condition for BOA is broken. However, as shown blow,
this approach could lead to the existence of a bound state wave function
$\phi $ which takes negligible value in this questionable region, such that
the discussion within BOA remains valid. Since the negative scattering
lengths do not support any two-body bound states, the existence of
such a three-body bound state in this region is of great interest.

\section{Three-body universal bound states in $1$D-$1$D-$3$D systems}

\begin{figure}[tbp]
\includegraphics[bb=35bp 285bp 583bp 541bp,clip,width=9cm]{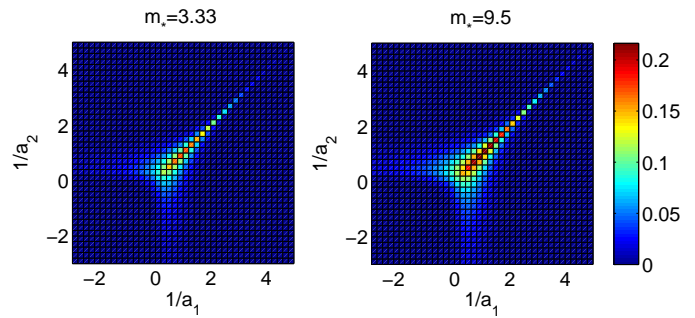}
\caption{(color online) The binding energy of the ground three-body
bound state in the $1$D-$1$D-$3$D system with reduced mass of
the heavy atoms $m_*=3.33$ and $9.5$. These values correspond to
the cases of ($A_{1}= A_2= ^{40}$K, $B= ^{6}$Li) and
($A_{1}=A_2=^{133}$Cs, $B=^{7}$Li), respectively.}
\label{be113}
\end{figure}

In the previous section, we have obtained the instantaneous bound-state wave function
$\psi$ of the light atom $B$ and the effective interaction potential $V_{\rm eff}(z_{12})$
between the two heavy atoms. We have shown that $V_{\rm eff}$
is most significant when the two-body scattering length is resonant with
the distance between the two $1$D tubes.
In this section we derive the wave functions and binding energies of the relevant
three-body bound states, and further confirm the observation of this new
resonance effect.

%As shown in Eq. (\ref{pf2}), the three-body
%bound state wave function $\Psi$ in such a system is given by $\Psi=\phi\psi$ under BOA, with $\phi$
%the bound-state solution of the Schr$\ddot{\rm o}$dinger equation
%(\ref{ef2}) with $z=z_1-z_2$. In this section we show our numerical
%results of the three-body ground state.

In Fig. \ref{be113}, the binding energy $E_{\rm 3b}$ of the ground
trimer state is plotted as a function of $1/a_1$ and
$1/a_2$ with heavy-atom reduced masses $m_*=3.33$ and $9.5$
in the natural unit. These values correspond to the cases of
($A_{1}= A_2= ^{40}$K, $B= ^{6}$Li) and ($A_{1}=A_2=^{133}$Cs, $B=^{7}$Li),
respectively. Here, the binding energy $E_{\rm 3b}$ is defined as the
energy gap between the three-body ground state $E$ and the threshold
of the effective interaction, i.e.
\begin{eqnarray}
E_{\rm 3b}=V_{\rm eff}(\infty) - E.
\end{eqnarray}

From Fig.~\ref{be113}, we notice that a three-body bound state
exists for a wide range of positive and negative scattering length
combinations, as discussed in the previous section. Nevertheless,
the binding energy reaches a peak value when the two scattering
lengths $a_1$ and $a_2$ are close with each other,
especially in the region around $a_1 \sim a_2 \sim 1$.
This observation is consistent with the expectation outlined
in the previous section, which shows that when $a_1 = a_2 = a$,
the effective potential well for $A_1$-$A_2$ interaction is deepest
as the scattering lengths are resonant with the distance between the two
1D tubes $a=1$.

\begin{figure}[tbp]
\includegraphics[bb=57bp 247bp 499bp 532bp,clip,width=7.5cm]{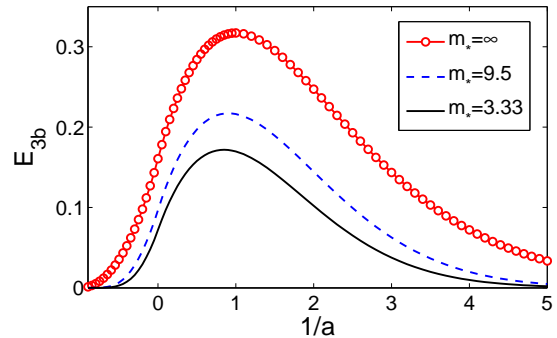}
\caption{(color online) The binding energy $E_{\rm 3b}$ of the ground
trimer state as a function of $1/a$ with $a_1=a_2=a$.
The reduced masses used in this plot are $m_*=3.33$
(black solid line), $9.5$ (blue dashed line) and $\infty$
(red solid line with open circles), respectively.}
\label{be113b}
\end{figure}

\begin{figure}[tbp]
\includegraphics[bb=44bp 272bp 457bp 549bp,clip,width=7.5cm]{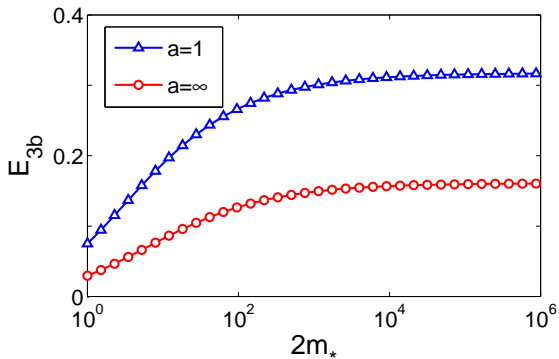}
\caption{(color online) The binding energy $E_{\rm 3b}$ of the ground
trimer state as a function of the reduced mass $m_*$
with $a_1=a_2=a=1$ (blue solid line with open triangle)
and $a_1=a_2=a=\infty$.}
\label{be113c}
\end{figure}

To further investigate the relationship between the binding energy and
the two-body scattering lengths, we focus on the case of $a_1 = a_2 = a$,
and illustrate in Fig.~\ref{be113b} the binding energy in terms of $1/a$ with
respect to different reduced masses $m_*$ of the two heavy atoms.
One significant feature of this result is that the resonant behavior is present
for all different reduced masses, i.e. the binding energy of the ground trimer
state reaches its maximum in the region around $a = 1$.
Besides, we also notice that for a given two-body scattering length,
the binding energy increases with reduced mass $m_*$, and
approaches to an asymptotic value in the limit $m_* \to \infty$.
This tendency is also confirmed by Fig.~\ref{be113c} where the
binding energies for $a_1=a_2=a=1$ and $a_1=a_2=a=\infty$ are
plotted as functions of the reduced mass $m_*$.

\begin{figure}[tbp]
\includegraphics[bb=52bp 252bp 488bp 540bp,clip,width=7.5cm]{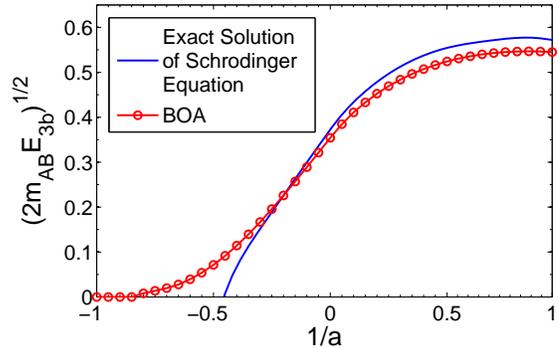}
\caption{(color online) The binding energy of the ground
three-body bound state in the 1D-1D-3D system with reduced
mass $m_*=3.33$ and scattering lengths $a_1=a_2=a$.
Here, we plot the results given by the BOA (red solid line with open circles)
and by the exact solution of the Schr$\ddot{\rm o}$dinger equation~\cite{nishida-11} (blue solid line). Notice that
the BOA can give reliable results provided that the binding energy of the
trimer state is away from the threshold.}
\label{be113d}
\end{figure}

The three-body bound states in the $1$D-$1$D-$3$D systems with
$a_1=a_2$ are also discussed in Ref.~\cite{nishida-11} within an
effective field theory or the exact solution of three-body
Schr$\ddot{\rm o}$dinger equation. In Fig. ~\ref{be113d},
we compare our BOA results of the ground trimer state
energy with the exact expression given by Ref.~\cite{nishida-11}
for $m_*=3.33$ and $a_1 =a_2 = a$. Notice that the BOA results
are very close to the exact solution around the resonance point $a=1$
for such a rather small mass ratio. This consistency suggests that
the BOA approach is reliable provided that the three-body bound
state energy is away from the threshold.

%%%%%%%%%%%%%
\section{Three-body universal bound states in $2$D-$2$D-$3$D systems}

The discussion on $1$D-$1$D-$3$D systems outlined in the previous section
can be directly generalized to other mixed-dimensional configurations. In this section
we consider a $2$D-$2$D-$3$D system [Fig. 1(b)] where the two heavy
atoms $A_{1,2}$ are trapped individually in two $2$D confinements,
localized in the planes of $x= \pm L/2$. The light atom $B$ is also
assumed to move freely in the $3$D space. we also
adopt the natural units with $\hbar = m_B = L =1$.

When the masses of $A_{1,2}$ is much larger than that of $B$, the system
can also be treated via BOA. The wave function $\Psi$ of the possible
three-body bound state also takes the factorized form as in Eq. (\ref{pf2}),
i.e., $\Psi=\phi\psi$ with $\psi$ the instantaneous bound state of the light atom $B$.
In this case, the instantaneous energy of $\psi$ serves as an effective $2$D
interaction between the two heavy atoms, and can be obtained by replacing
the argument $z_1-z_2$ in $V_{\rm eff}(z_1-z_2)$ with
$\rho=\sqrt{(y_1-y_2)^2+(z_1-z_2)^2}$.
Following the same procedure as outlined in Sec. II, we can show that
in the case of $a_1=a_2=a$, the depth of the $2$D effective potential also takes
its maximal value when $a=1$ in the natural unit. This observation indicates
that the resonance phenomenon also exists in the $2$D-$2$D-$3$D configuration.

Notice that the 2D-2D-3D geometry is invariable under a rotation along the $x$-axis.
This $\rm{SO(2)}$ symmetry thus leads to the conservation of the $x$-component
angular momentum of $A_1$-$A_2$ relative motion. Therefore, the wave
function $\phi$ in the three-body bound state $\Psi$ can be expressed as
\begin{eqnarray}
\phi= \sum_{\ell} \phi_\ell(\rho)e^{i \ell \theta},
\end{eqnarray}
where $\tan\theta=(z_1-z_2)/(y_1-y_2)$ is the polar angle of $A_{1,2}$
relative motion in the $y$-$z$ plane,
and the radial wave function $\phi_\ell(\rho)$ satisfies the
$2$D Schr\"{o}dinger equation
\begin{eqnarray}
&&\left[-\frac{1}{2 m_*}\left(\frac{d^2}{d\rho^2}+\frac{1}{\rho}\frac{d}{d\rho}
-\frac{\ell^2}{\rho^2}\right)
+V_{\rm eff}(\rho)\right]\phi_\ell(\rho)
\nonumber \\
&& \hspace{4cm}
=E_l\phi_\ell(\rho).  \notag \\
\label{4.phi}
\end{eqnarray}
Here, the quantum number $\ell=0,\pm 1,\pm 2,...$ indicates
the relative angular momentum of $A_{1,2}$ along the $x$-direction.
The pure ground state of the system occurs in the channel $\ell=0$.

\begin{figure}[tbp]
\includegraphics[bb=5bp 274bp 539bp 530bp,clip,width=9cm]{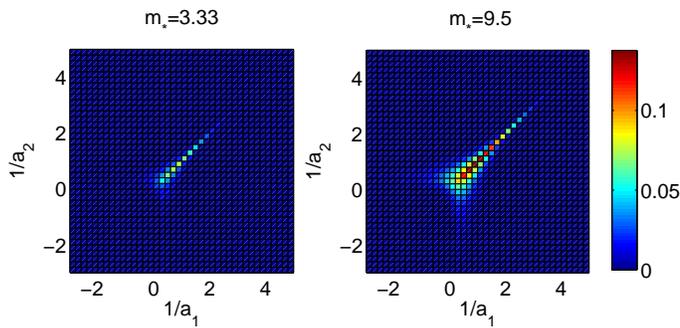}
\caption{(color online) The binding energy of the ground three-body
bound state in the $2$D-$2$D-$3$D geometry with reduced mass
of the heavy atoms $m_*=3.33$ and $9.5$.}
\label{be223}
\end{figure}

\begin{figure}[tbp]
\includegraphics[bb=35bp 263bp 489bp 559bp,clip,width=7cm]{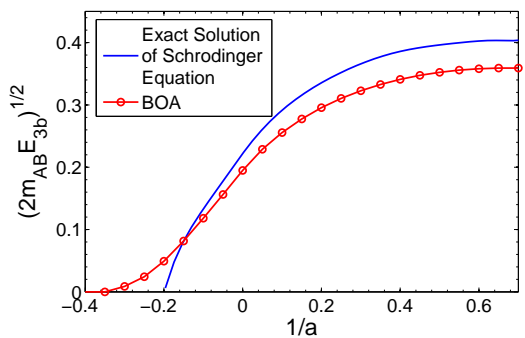}
\caption{(color online) The binding energy of the ground three-body
bound state in 2D-2D-3D geometry with reduced mass $m_*=3.33$ and
scattering lengths $a_1=a_2=a$. Here, we plot the results given by
the BOA (red solid line with open circles) and by an
effective field theory~\cite{nishida-10} (blue solid line), and find
good agreement provided that the binding energy is away from
the threshold.}
\label{fig.4.E1}
\end{figure}

The radial equation (\ref{4.phi}) can be solved numerically as in the
1D-1D-3D case. For the ground zero-angular momentum channel $\ell =0$,
we also find three-body bound states with reduced mass $m_* = 3.33$
and 9.5. The binding energy of the ground trimer state
is illustrated in Fig. \ref{be223} in terms of $1/a_1$ and $1/a_2$.
Notice that the binding energy is significantly amplified in the
parameter region $a_1 \sim a_2$, and reaches its maximum when
the scattering lengths are resonant with the 2D surfaces spacing
$a_1 \sim a_2 \sim 1$. Besides, the binding energy also increases with
reduced mass $m_*$ of the two heavy atoms.
In Fig.~\ref{fig.4.E1} we also compare the BOA results
with the exact expression~\cite{nishida-10} for the case of $m_*=3.33$
and $a_1=a_2=a$, and find good agreement when the trimer binding
energy is away from the threshold. All these features are analogous with
the case of $1$D-$1$D-$3$D geometry.

\section{Four-body universal bound states in 1D-1D-1D-3D systems}

From the discussion in the previous sections, we notice that the
BOA works well throughout a wide range of scattering length
for a fairly small mass ratio of about $6$, provided that the binding energy
of the bound trimer state is away from the threshold. This observation
suggests that this approach can be directly applied to mixed dimensional
systems with more than three atoms, and to give reliable results for few-body
bound state energy when it is sizable. In this section, we consider as an example
the 1D-1D-1D-3D system with three heavy atoms $A_1$, $A_2$, and $A_3$
trapped individually in parallel 1D tubes and a single light atom $B$ moving freely in 3D.

\begin{figure}[tbp]
\begin{centering}
\includegraphics[bb=120bp 431bp 467bp 775bp,clip,width=7cm]{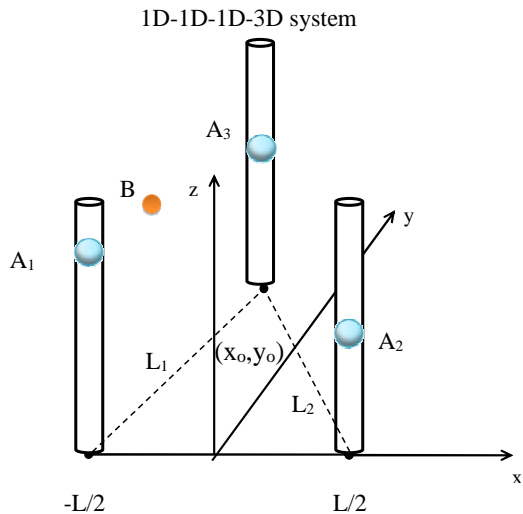}
\caption{(color online) The 1D-1D-1D-3D system with three heavy atoms
$A_1, A_2$ and $A_3$ confined in three 1D tubes and the light atom $B$ moving freely in 3D.}
\end{centering}
\label{fig1113}
\end{figure}

We consider the configuration of three 1D tubes
arranged along the $z$ direction, and intersect with the
$x$-$y$ plane at $(x=\pm L/2, y=0)$ and $(x=x_0, y=y_0)$,
as shown schematically in Fig.~\ref{fig1113}.
The three intersection points form a triangle in the $x$-$y$ plane.
Since the system properties are invariant under different
length scales, we assume that $L$ is the shortest side of the triangle,
and use it as the length unit $L=1$ in the following discussion.

The quantum states of such a system can be described by the wave function
$\Psi (\vec{r}_{B};z_{1},z_{2}, z_{3})$, where $z_i$ is the $z$-coordinate of the
heavy atom $A_i$, and $\vec{r}_{B}$ is the coordinate of the light atom $B$.
Within the BOA, the wave function $\Psi$ can be separated as
\begin{eqnarray}
\label{eqn:4body-BOA}
\Psi(\vec{r}_{B};z_{1},z_{2}, z_{3}) = \phi(z_{1},z_{2}, z_{3}) \psi(\vec{r}_{B};z_{1},z_{2}, z_{3}).
\end{eqnarray}
Here, $\psi$ is the wave function of the instantaneous bound state of the light atom,
which is given by the Schr{\"o}dinger equation
\begin{eqnarray}
\label{eqn:4body-SE}
- \frac{1}{2}\nabla_B^2 \psi(\vec{r}_{B};z_{1},z_{2}, z_{3})
=
V_{\rm eff} (z_{1},z_{2}, z_{3}) \psi(\vec{r}_{B};z_{1},z_{2}, z_{3})
\nonumber\\
\end{eqnarray}
with the Bethe-Peierls boundary conditions
\begin{eqnarray}
\label{eqn:4body-BP}
\Psi(r_{iB} \to 0) \propto \left( 1- \frac{a_i}{r_{iB}}\right) +\mathcal{O}(r_{iB}).
\end{eqnarray}
Here, $r_{iB}$ and $a_i$ are the distance and mixed-dimensional scattering length
between the atoms $A_i$ and $B$, respectively.

%This eigen-equation
%determines the effective potential $V_{\rm eff}$ among the heavy atoms,
%and should be solved for a given set of $(z_{1},z_{2}, z_{3})$.

The ground state of $\psi$ can be obtained by solving the
eigen-equation (\ref{eqn:4body-SE}) for a give set of $(z_{1},z_{2}, z_{3})$,
which takes the form
\begin{eqnarray}
\label{eqn:4body-psi}
\psi(\vec{r}_{B};z_{1},z_{2}, z_{3}) &=&
\frac{e^{-\kappa r_{1B}}}{r_{1B}} + c_2 \frac{e^{-\kappa r_{2B}}}{r_{2B}}
+ c_2 \frac{e^{-\kappa r_{2B}}}{r_{2B}},
\nonumber \\
V_{\rm eff} (z_1, z_2, z_3) &=& -\frac{\kappa^2}{2}.
\end{eqnarray}
The parameter $\kappa$ is determined by the boundary conditions (\ref{eqn:4body-BP}),
leading to
\begin{eqnarray}
\label{eqn:4body-kappa}
\kappa - c_2 \frac{e^{-\kappa r_{12}}}{r_{12}} - c_3 \frac{e^{-\kappa r_{13}}}{r_{13}}
&=& \frac{1}{a_1};
\nonumber \\
- \frac{e^{-\kappa r_{12}}}{r_{12}} + c_2 \kappa - c_3 \frac{e^{-\kappa r_{23}}}{r_{23}}
&=& \frac{c_2}{a_2};
\nonumber \\
- \frac{e^{-\kappa r_{13}}}{r_{13}} - c_2 \frac{e^{-\kappa r_{23}}}{r_{23}} + c_3 \kappa
&=& \frac{c_3}{a_3}.
\end{eqnarray}
A numerical solution of these equations hence gives the effective potential $V_{\rm eff}$
among the three heavy atoms. By imposing the potential to the Schr{\"o}dinger equation
\begin{eqnarray}
\label{eqn:4body-SE2}
&&\left[ - \sum_{i = 1,2,3} \frac{1}{2m_i} \frac{\partial^2}{\partial z_i^2} + V_{\rm eff} (z_{1},z_{2}, z_{3})\right]
\phi(z_{1},z_{2}, z_{3})
\nonumber \\
&& \hspace{4cm}
= E \phi(z_{1},z_{2}, z_{3}),
\end{eqnarray}
we can obtain the energy $E$ for the four-body bound states.

From now on, we focus on the special case of $a_1 = a_2 = a_3 = a$
and $m_1 = m_2 =m_3 =m$, that is the scattering lengths and the masses
of the three heavy atoms are all the same. This is also the most relevant
case for experiments, where atoms trapped in low dimensional traps
are of the same species. Since the system is translationally invariant
along the $z$-direction, we define a new set of coordinates
\begin{eqnarray}
X &=& z_1 - z_2,
\nonumber \\
Y &=& z_3 - \frac{z_1 + z_2}{2},
\end{eqnarray}
and calculate the effective potential $V_{\rm eff} (X,Y)$ in these new variables
\begin{eqnarray}
\label{eqn:4body-Ueff}
V_{\rm eff}(X,Y) = U(a; X, Y) - \frac{1}{2a^2},\label{bigu}
\end{eqnarray}
where $U(a; X,Y)$ is the regularized part.

\begin{figure}[tbp]
\includegraphics[bb=57bp 236bp 569bp 568bp,clip,width=7cm]{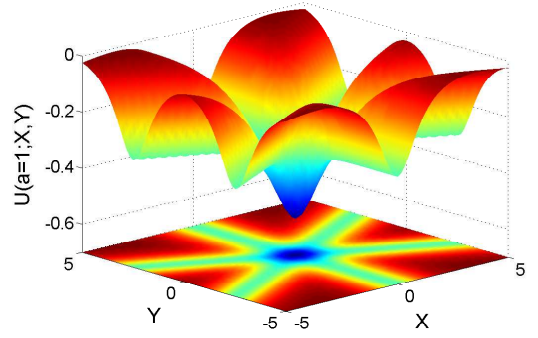}
\caption{(color online) The regularized effective
potential $U(a;X,Y)$ for two body scattering length
$a_1=a_2=a_3=a=1$ in the 1D-1D-1D-3D system 
with equilateral triangle configuration.}
\label{fig:4body-potential}
\end{figure}

We first consider the special geometry where the three 1D tubes
are arranged equidistantly to form an equilateral triangle in the $x$-$y$ plane
(i.e., $x_0=0$ and $y_0=\sqrt{3}/2$).
%{\color{blue}This configuration can be obtained by setting three pairs
%of laser beams 120$^\circ$ apart to form a triangular optical lattice.}
In Fig.~\ref{fig:4body-potential}, we show the regularized effective
potential $U(a;X,Y)$ for scattering length $a=1$. Notice that the effective
potential acquires its global minimum at $(X=0, Y=0)$ or $z_1=z_2=z_3$,
that is the three atoms are staying in a surface perpendicular to the 1D tubes
and forming an equilateral triangle. Besides, we also observe three
energy potential valleys, which correspond to the cases where the distance
between two of the three atoms equals to $1$.

\begin{figure}[tbp]
\includegraphics[bb=66bp 236bp 493bp 535bp,clip,width=7cm]{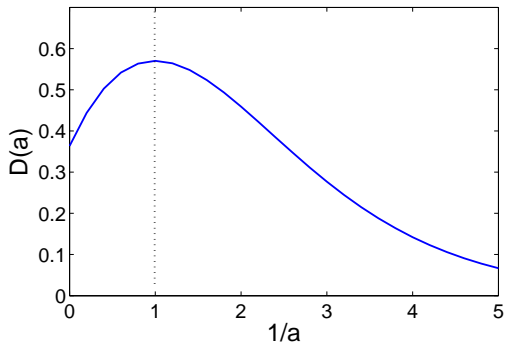}
\caption{(color online) The depth $D(a)$ of the regularized part $U(a; X,Y)$
of the effective interaction. In this plot, we consider the case of $a_1=a_2=a_3=a$
in the 1D-1D-1D-3D system with equilateral triangle configuration.
Notice that $D(a)$ takes maximum value at
the resonance point of $a=1$.}
\label{fig:4body-Da}
\end{figure}

The same phenomenon can also be observed for
other values of scattering length $a \neq 1$. In fact, the effective potential
$U(a; X,Y)$ always reaches its minimum at $(X=0, Y=0)$.
However, the potential is deepest only when the scattering length $a=1$.
In Fig.~\ref{fig:4body-Da}, we show the depth of the effective potential
well as a function of $a$, which reaches its maximum at $a = 1$.
This result suggests that the resonance we observed in three-body problems
as discussed above also occurs in the four-body system.

With the knowledge of the effective potential, we can numerically solve the
Schr{\"o}dinger equation (\ref{eqn:4body-SE2}) to obtain the eigenenergies of
four-body bound states. In the new set of variables $X$ and $Y$, this equation
can be rewritten as
\begin{eqnarray}
\label{eqn:4body-SE3}
&&\left[ - \frac{1}{m} \frac{\partial^2}{\partial X^2}
- \frac{3}{4m} \frac{\partial^2}{\partial Y^2}
+
V_{\rm eff} (X,Y)\right]
\phi(X,Y)
\nonumber \\
&& \hspace{4cm}
= E \phi(X,Y),
\end{eqnarray}
where $\phi(X,Y)$ is the wave function of the heavy atoms.
As in the three-body calculation, the binding energy of the tetramer states is
defined as the difference between the eigenenergy $E$ and the effective
potential energy for $X \to \infty$ and $Y \to \infty$,
\begin{eqnarray}
\label{eqn:4body-bindingE}
E_{\rm 4b} = V_{\rm eff}(\infty,\infty) - E.
\end{eqnarray}

\begin{figure}[tbp]
\includegraphics[bb=45bp 274bp 437bp 512bp,clip,width=8cm]{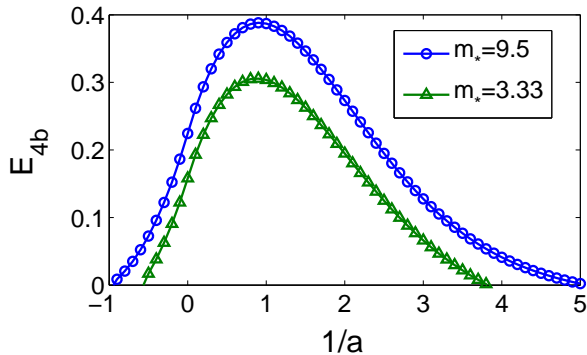}
\caption{(color online) The binding energy $E_{\rm 4b}$ of the ground
four-body bound state in the 1D-1D-1D-3D system  with equilateral triangle configuration.
for different values of $a$ with reduced mass
$m_*=9.5$ (blue solid line with circles) and 3.33 (green solid line with triangle).}
\label{fig:4body-bindingE1}
\end{figure}

The binding energy of the ground four-body bound state for different
values of $a$ is plotted in Fig.~\ref{fig:4body-bindingE1}, where we consider
two mass ratios as in the previous discussion. Notice that the binding energy
reaches its maximum near $a = 1$, as we expected from the effective potential.
This result confirms the appearance of the resonance phenomenon
in the four-body system.

\begin{figure}[tbp]
\includegraphics[bb=29bp 239bp 437bp 555bp,clip,width=7cm]{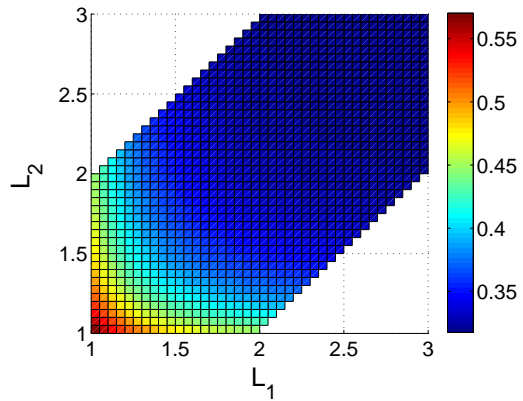}
\caption{(color online) The depth of the effective potential
$U(a;X,Y)$ in the 1D-1D-1D-3D system with $a_1=a_2=a_3=1$
and the inter-tube distances $L_1=1$ and $L_{1,2}$ defined in Fig. 10.}
\label{fig:4body-geometry}
\end{figure}

Up to now, we consider only a special configuration of 1D-1D-1D-3D geometry
where the three 1D tubes form an equilateral triangle, and observe a resonance
phenomena for tetramer binding energy as the scattering length gets close to
the mutual distance between 1D tubes. An intuitive expectation is that this most
symmetric configuration should be the case of maximal resonance, for the scattering
length can be resonant with any two of the three atoms. In order to demonstrate
this idea, we consider general configurations of the three 1D tubes, such that
they form a triangle of arbitrary shape with three sides $L=1$, $L_1$ and $L_2$
(see Fig. 10). For the system properties are invariant as scaled
with length, we assume $L=1$ to be the shortest side of the triangle. We further 
take the scattering lengths $a_1=a_2=a_3=1$.
In Fig.~\ref{fig:4body-geometry}, we show the depth of the effective potential
$U(a;X,Y)$ for arbitrary arrangement of the three 1D tubes. It is
clearly shown that, the depth of the effective potential takes its maximum value
when $L_1=L_2=1$ or the $1$D tubes have the configuration of equilateral triangle.
This is consistent with our expectation that maximal resonance
appears in this most symmetric configuration.

\section{BOA for many-body problems in mixed-dimensional systems}

In the previous sections, we study the three-body and four-body
bound states in mixed-dimensional systems within the BOA.
Now we generalize this approach to mixed-dimensional problems
with arbitrary $N$ heavy atoms trapped individually in 1D or 2D confinements,
while a single light atoms moving freely in the 3D space.
In such a configuration, the wave function of the possible few-body
bound states takes the form
\begin{eqnarray}
\Psi(\vec r_B;\vec s)=\phi(\vec s)\psi(\vec r_B;\vec s),
\end{eqnarray}
where $\vec s=(\vec{r}_1,\vec{r}_2,..,\vec{r}_N)$ are the 1D or 2D coordinates
of the heavy atoms  $A_1,A_2,...,A_N$,
and ${\vec r}_B$ is the coordinate of the light atom $B$.
As in the previous sections, $\psi(\vec r_B;\vec s)$ is the wave
function of the instantaneous bound state of the light atom,
which is determined by the Schr\"{o}dinger equation
\begin{eqnarray}
-\frac{1}{2}\nabla^2_B\psi(\vec r_B;\vec s)
=V_{\rm eff}(\vec s)\psi(\vec r_B;\vec s)
\label{5.psi}
\end{eqnarray}
with Bethe-Peierls boundary conditions
\begin{eqnarray}
\psi(r_{iB}\rightarrow0)\propto\left(1-\frac{a_i}{r_{iB}}\right)
+\mathcal{O}(r_{iB}). \label{5.bethe}
\end{eqnarray}
Here, $r_{iB}$ is the distance between the atoms $A_i$ and $B$.

By solving Eq. (\ref{5.psi}), we obtain the general form of
the instantaneous bound state
\begin{eqnarray}
\psi(\vec r_B;\vec s)=\frac{e^{-\kappa r_{1B}}}{r_{1B}}+\sum_{i=2}^N
c_i\frac{e^{-\kappa r_{iB}}}{r_{iB}},
\label{5.psi2}
\end{eqnarray}
where the value of $\kappa$ and the coefficients $c_i$ are given
by the equations
\begin{eqnarray}
\frac{1}{a_1} &=&
\kappa -\sum_{j=2}^Nc_i\frac{e^{-\kappa r_{iB}}}{r_{iB}}; \\
\frac{c_l}{a_l} &=& \kappa c_l-\frac{e^{-\kappa r_{lB}}}{r_{lB}}
-\sum_{i=2,i\neq l}^Nc_i\frac{e^{-\kappa r_{iB}}}{r_{iB}}.
\end{eqnarray}
From the equations above, we can solve for the value of $\kappa$
in terms of the coordinate $\vec{s}$ of the heavy atoms, and then obtain
the instantaneous wave function of $\psi(\vec r_B;\vec s)$ and the effective
interaction among the heavy atoms
\begin{eqnarray}
V_{\rm eff}(\vec s)=-\frac{\kappa^2}{2}.
\end{eqnarray}
Finally, the heavy-atoms wave function $\phi(\vec s)$ of the few-body
bound state is given by
\begin{eqnarray}
\left[-\sum_{i=1}^N\frac{1}{2m_i}\nabla^2_i+V_{\rm eff}(\vec s)\right]%
\phi(\vec s)=E\phi(\vec s)
\end{eqnarray}
with $m_i$ the mass of the heavy atom $A_i$.

\section{Conclusion}

In this manuscript we show our BOA-based results on the stable
three-body or four-body bound states in mixed dimensional systems with $N \ge 2$ heavy atoms
individually trapped in different 1D or 2D confinements, while a single light atom
moving freely in the 3D space.  The BOA approach can
provide a clear physical picture with a well-defined effective interaction among the heavy atoms.
We show that in mixed dimensions,
the three-body or four-body bound states can occur within a broad range of two-body scattering
lengths, as the Efimov states in 3D. Nevertheless, the binding energy of the
ground bound state reaches its maximum value when the two-body scattering
length gets close to the distance between the low-dimensional traps.
This is due to a new resonance phenomenon in mixed dimensions,
where the effective interaction among the heavy atoms
acquires a deepest potential well under the resonant condition.
The feasibility of this BOA approach is then confirmed by a direct
comparison with exact results in 1D-1D-3D and 2D-2D-3D configurations,
hence suggests a possible extension in the problems with more than
three atoms in mixed dimensions.

%Before finalizing this manuscript, we notice a manuscript has been
%recently posted~\cite{nishida-11}, where the stable three-body bound
%states in 1D-1D-3D and 2D-2D-3D geometries have been analyzed.
%However, the resonance effect we find here is not discussed.

\begin{acknowledgments}
This work is supported by National Natural Science Foundation of China
(11074305, 10904172), the Fundamental Research Funds for the
Central Universities, and the Research Funds of Renmin University
of China (10XNL016). WZ would also like to thank the China Postdoctoral
Science Foundation and NCET Program for support.
\end{acknowledgments}

\appendix%\appendixpage
\addcontentsline{toc}{section}{Appendices}\markboth{APPENDICES}{}
\begin{subappendices}

\section{The Bethe-Peierls Boundary Condition for BOA in
Mixed-Dimensional Systems}

In this appendix, we derive the Bethe-Peierls boundary condition
[e.g., Eqs. (\ref{c1}), (\ref{c2}), (\ref{eqn:4body-BP}) and (\ref{5.bethe})]
used in the Born-Oppenheimer approach for the mixed dimensional systems.
For simplicity, here we consider the system with one heavy atom $A$
confined in a $1$D trap which is arranged along the $z$-axis,
plus a light atom $B$ moving freely in $3$D. The generalization
to other cases is straightforward.

The expression of Bethe-Peierls boundary condition should be
derived from the asymptotic behavior of the two-body wave functions.
As a first-principle discussion, we first take into account the $3$D motions
of both atoms $A$ and $B$, and then reduce our result in the
mixed-dimensional model where only the motion along the $z$ direction is
considered for atom $A$. The total Hamiltonian of the two atoms is given by
\begin{eqnarray}
H_{AB}=T_{Az}+T_{A\perp }+V_{A\perp }+T_{B}+V_{AB}\left( r_{AB}\right) .
\label{bobph2b}
\end{eqnarray}%
Here, the kinetic energy of atom $A$ along the $z$ direction is given by
\begin{eqnarray}
T_{Az}=-\frac{1}{2m_{A}}\frac{\partial ^{2}}{\partial z_{A}^{2}}
\end{eqnarray}%
with $m_{A}$ the mass of atom $A$, and $\vec{r}_{i=A,B}=\left(
x_{i},y_{i},z_{i}\right) $ the coordinate of the corresponding atoms.
The transverse kinetic energy $T_{A\perp }$ of atom $A$ and the
total kinetic energy $T_{B}$ of atom $B$ are defined as
\begin{eqnarray}
T_{A\perp } &=&-\frac{1}{2m_{A}}\left( \frac{\partial ^{2}}{\partial
x_{A}^{2}}+\frac{\partial ^{2}}{\partial y_{A}^{2}}\right); \\
T_{B} &=&-\frac{1}{2}\nabla_{B}^2.
\end{eqnarray}
Here we use the natural unit $\hbar=m_B=1$. In the Hamiltonian (\ref{bobph2b})
we also have the transverse harmonic potential
\begin{eqnarray}
V_{A\perp }=\frac{m_{A}\omega _{\perp }^{2}}{2}\left(
x_{A}^{2}+y_{A}^{2}\right)
\end{eqnarray}%
with frequency $\omega _{\perp }$, and the
atom-atom interaction potential $V_{AB}\left( r_{AB}\right) $
which is a function of the distance between the two particles
$r_{AB}=\left\vert \vec{r}_{A}-\vec{r}_{B}\right\vert$.
We further denote the effective range of the interaction potential as $r_{\ast }$,
such that we have $V_{AB}\left( r_{AB}\right) \approx 0$ in the region of
$r_{AB} \gg r_{\ast }$.

When the confinement of the heavy atom $A$ is strong, the transverse motion
of atom $A$ in the $x$-$y$ plane is much more rapid than its motion along the
$z$ direction. Therefore, we need to consider both the position $\vec{r}_{B}$
of the light atom $B$ and the transverse coordinates
$\left(x_{A},y_{A}\right) $ of the heavy atom $A$ as fast degrees of freedom.
Only the longitudinal coordinate $z_{A}$ of atom $A$ is treated as the slow
variable.

Within the BOA, the total wave function of the system takes the form%
\begin{eqnarray}
\Psi \left( \vec{r}_{A},\vec{r}_{B}\right) =\phi \left( z_{A}\right) \psi (%
\vec{r}_{B},x_{A},y_{A};z_{A}),  \label{bobpbigpsi}
\end{eqnarray}%
where $\psi (\vec{r}_{B},x_{A},y_{A};z_{A})$ is given by the eigen-equation%
\begin{eqnarray}
H_{F}\left( z_{A}\right) \psi (\vec{r}_{B},x_{A},y_{A};z_{A})=E\left(
z_{A}\right) \psi (\vec{r}_{B},x_{A},y_{A};z_{A})\nonumber\\  \label{bobpee}
\end{eqnarray}%
of the Hamiltonian
\begin{eqnarray}
H_{F}\left( z_{A}\right) =T_{A\perp }+V_{A\perp }+T_{B}+V_{AB}\left(
r_{AB}\right)   \label{bobphf}
\end{eqnarray}%
with fixed values of $z_{A}$. To solve Eq. (\ref{bobpee}), we expand the solution $\psi $
with eigen-states of the transverse Hamiltonian $T_{A\perp }+V_{A\perp }$ of atom $%
A$
\begin{eqnarray}
\psi (\vec{r}_{B},x_{A},y_{A};z_{A})=\sum_{n=0}^{\infty} \phi _{n}
\left(x_{A},y_{A}\right) \psi _{n}\left( \vec{r}_{B};z_{A}\right) .  \label{psi}
\end{eqnarray}
Here, $\phi _{n}\left( x_{A},y_{A}\right) $ is the $n^{\rm th}$ eigen-state of
$T_{A\perp }+V_{A\perp }$. Considering the translational symmetry along the
$z$-axis, we take $z_{A}=0$, and the relevant wave function
$\psi _{n}\left( \vec{r}_{B};0\right) $ of the light atom $B$ is given by
\begin{eqnarray}
&&\left[ T_{B}+\left( n+1\right) \omega _{\perp }\right] \psi_{n}
+
\left[ \sum_{m}V_{nm}\left( \vec{r}_{B}\right) \psi_{m} \right]
\nonumber \\
&&\hspace{4cm}
=
E\left( 0\right) \psi _{n}. \label{bobpe22}
\end{eqnarray}
Here, the matrix element of the interaction potential takes the form
\begin{eqnarray}
V_{nm}\left( \vec{r}_{B}\right) &=&
\int dx_{A}dy_{A}\phi _{n}^{\ast }\left(x_{A},y_{A}\right)
\nonumber \\
&& \hspace{0.5cm}
\times V_{AB}\left( r_{AB}\right) \phi _{m}\left(
x_{A},y_{A}\right) . \label{bobpvmn}
\end{eqnarray}
Therefore, the eigen-equation (\ref{bobpee}) or (\ref{bobpe22}) can be solved via a multi-channel
scattering theory of atom $B$, with the transverse states
$\phi _{n}\left( \vec{r}_{B};z_{A}\right) $ of atom $A$ serving as the
scattering channels. In the low-energy case with
$\omega _{\perp }<E<2 \omega _{\perp }$,
the ground channel with the transverse state $\phi _{0}\left(x_{A},y_{A}\right) $
assumes the only open channel.

Now we consider the asymptotic behavior of the wave function in the
long-distance limit with $\left\vert \vec{r}_{B}\right\vert \gg (r_{\ast},l_{\perp })$,
where $l_{\perp }=\sqrt{1/\left( m_{A}\omega _{\perp }\right) }$ denotes
the characteristic length of the transverse confinement. In this region,
the mutual distance $r_{AB}$ between the two atoms would be much larger
than the effective range $r_{\ast }$ of the interaction, such that we can
neglect the term $V_{AB}$ in Eq. (\ref{bobphf}).
According to the scattering theory, in such a region the wave function $\psi
_{n}\left( \vec{r}_{B};0\right) $ in the close channels with $n>0$
decays exponentially with $\left\vert \vec{r}_{B}\right\vert $, and can be safely neglected.
The wave function $\psi _{0}\left( \vec{r}_{B};0\right) $
in the open channel takes the form
\begin{eqnarray}
\psi _{0}\left( \vec{r}_{B};0\right)  &\sim &\sum_{l=0}^{\infty
}\sum_{m=-l}^{l}C_{l,m}\frac{Y_{lm}\left( \theta _{B},\phi _{B}\right) }{%
k\left\vert \vec{r}_{B}\right\vert }  \nonumber  \label{bobppsi0a} \\
&&\hspace{-1.5cm}
\times \left( \hat{\jmath}_{l}\left( k\left\vert \vec{r}_{B}\right\vert
\right) +kf_{l,m}\left( k\right) \hat{h}_{l}^{\left( +\right) }\left(
k\left\vert \vec{r}_{B}\right\vert \right) \right),
\end{eqnarray}%
where $k=\sqrt{2\left( E- \omega _{\perp }\right) }$,
$Y_{lm}\left( \theta ,\phi \right) $ are the spherical harmonic functions of the
azimuth angles $\left( \theta _{B},\phi _{B}\right) $ of $\vec{r}_{B}$,
$\hat{\jmath}_{l}\left( z\right) $ is the Riccati-Bessel function,
and $\hat{h}_{l}^{\left( +\right) }\left( z\right) $ is the Riccati-Hankel function.
The coefficients $C_{l,m}$ are given by the boundary condition,
while the scattering amplitudes $f_{l,m}\left( k\right) $ are determined by the
effective potential $V_{nm}\left( \vec{r}_{B}\right) $ defined in (\ref{bobpvmn}).
In the low-energy case with small $k$, we can neglect all the
high-partial wave scattering amplitudes $f_{l,m}\left( k\right) $ with $l>0$%
, and approximate the $s$-wave scattering amplitude $f_{0,0}\left( k\right) $
with $f_{0,0}\left( k=0\right) $. Then the long-distance behavior of wave
function $\psi $ becomes%
\begin{eqnarray}
&&\psi (\vec{r}_{B},x_{A},y_{A};0) \simeq \phi _{0}\left( x_{A},y_{A}\right)
\psi _{0}\left( \vec{r}_{B};0\right)   \nonumber \\
&&\sim \phi _{0}\left( x_{A},y_{A}\right) \left[ \frac{1}{k\left\vert \vec{r}%
_{B}\right\vert }\left( \sin \left( k\left\vert \vec{r}_{B}\right\vert
\right) -ka_{AB}e^{ik\left\vert \vec{r}_{B}\right\vert }\right) \right.
\nonumber \\
&&\left. +\sum_{l=1}^{\infty }\sum_{m=-l}^{l}C_{l,m}\frac{Y_{lm}\left(
\theta _{B},\phi _{B}\right) }{k\left\vert \vec{r}_{B}\right\vert }\hat{%
\jmath}_{l}\left( k\left\vert \vec{r}_{B}\right\vert \right) \right]
\label{bobppsi0b}
\end{eqnarray}%
with the scattering length $a_{AB}$ defined as
\begin{eqnarray}
a_{AB}=-f_{0,0}\left( k=0\right) .  \label{bobpaab}
\end{eqnarray}

The expression (\ref{bobppsi0b}) implies that in the \textquotedblleft
intermediate" region of
\begin{eqnarray}
\left[ r_{\ast },l_{\perp } \right] \ll \left\vert \vec{r}_{B}\right\vert \ll \frac1k,
\end{eqnarray}
the behavior of $\psi $ takes the form of
\begin{eqnarray}
\psi (\vec{r}_{B},x_{A},y_{A};0)\sim \phi _{0}\left( x_{A},y_{A}\right)
\left( 1-\frac{a_{AB}}{\left\vert \vec{r}_{B}\right\vert }\right) .
\end{eqnarray}%
Therefore, we can replace the real interaction potential $V_{AB}\left(
r_{AB}\right) $ in (\ref{bobph2b}) with a Bethe-Peierls-type boundary
condition%
\begin{eqnarray}
\lim_{\left\vert \vec{r}_{B}\right\vert \rightarrow 0}\psi (\vec{r}%
_{B},x_{A},y_{A};0)\propto \phi _{0}\left( x_{A},y_{A}\right) \left( 1-\frac{%
a_{AB}}{\left\vert \vec{r}_{B}\right\vert }\right) .  \label{bobpbobp}
\end{eqnarray}%
Under this boundary condition, the solution of the eigen-equation%
\begin{eqnarray}
&&\left[ T_{A\perp }+V_{A\perp }+T_{B}\right] \psi (\vec{r}%
_{B},x_{A},y_{A};0)
\nonumber\\
&& \hspace{3cm}
=E \psi (\vec{r}_{B},x_{A},y_{A};0)
\end{eqnarray}%
takes the form of Eq. (\ref{bobppsi0b}) for all $\left\vert \vec{r}%
_{B}\right\vert \neq 0$, and becomes a reasonable approximation
for the solution of (\ref{bobpee}).

In this reduced mixed-dimensional model, the transverse
coordinates $\left( x_{A},y_{A}\right) $ of the heavy atom $A$ is taken to be
fixed values of $\left( 0,0\right) $. Together with the assumption $z_{A}=0$%
, we have%
\begin{eqnarray}
\left\vert \vec{r}_{B}\right\vert =r_{AB},
\end{eqnarray}%
and then the boundary condition (\ref{bobpbobp}) can be expressed as%
\begin{eqnarray}
\lim_{r_{AB}\rightarrow 0}\psi \left( \vec{r}_{B};0\right) \propto \left( 1-%
\frac{a_{AB}}{r_{AB}}\right) .  \label{bobpnb}
\end{eqnarray}%
Here, $\psi \left( \vec{r}_{B};0\right) $ is the wave function of the light
atom $B$ with the position of atom $A$ fixed at $z_{A}=0$. For non-zero
$z_{A}$, the condition (\ref{bobpnb}) can be generalized to
\begin{eqnarray}
\lim_{r_{AB}\rightarrow 0}\psi \left( \vec{r}_{B};z_{A}\right) \propto
\left( 1-\frac{a_{AB}}{r_{AB}}\right) .  \label{bobpnb3}
\end{eqnarray}%
That is the Bethe-Peierls boundary condition used in the BOA discussed
in the main text of this manuscript.

We notice that there is another type of Bethe-Peierls boundary condition
as discussed in Ref.~\cite{nishida-08,nishida-11}, where the total
wave function $\Psi$ of the reduced mixed-dimensional two-body problem
is assumed to satisfy the condition
\begin{eqnarray}
\lim_{D_{AB}\rightarrow 0}\Psi \propto \left( 1-\frac{a_{\rm eff}}{D_{AB}}%
\right)   \label{bobp2}
\end{eqnarray}%
with
\begin{eqnarray}
D_{AB}=\sqrt{x_{B}^{2}+y_{B}^{2}+\frac{m_{A}+1}{m_{A}}\left(
z_{A}-z_{B}\right) ^{2}}.
\end{eqnarray}%
This condition is slightly different from our result of Eq. (\ref{bobpnb3}).
The difference can be understood by noticing that when solving for
the wave function of atom $B$ under BOA, we fix the position of the heavy atom
$A$, such that the relevant Bethe-Peierls boundary condition (\ref{bobp2})
becomes isotropic. It is pointed out that, in the limit of $m_{A} \gg 1$,
the condition (\ref{bobp2}) approaches to (\ref{bobpnb3}) and we have
$a_{AB}=a_{\rm eff}$. Therefore, we approximate $a_{AB}$ as $a_{\rm eff}$
when comparing the BOA results with the effective field theory~\cite{nishida-11}.

It is straightforward to generalize the discussion above to more general
cases with $N$ heavy atoms $A_{1},...,A_{N}$ individually confined in $N$
low-dimensional traps, and one light atom $B$ moving freely in $3$D.
In that case, we can fix the positions of the heavy atoms under BOA,
and use the Bethe-Peierls boundary condition
\begin{eqnarray}
\lim_{r_{iB}\rightarrow 0}\psi \left( \vec{r}_{A},\vec{r}_{B}\right) \propto
\left( 1-\frac{a_{iB}}{r_{iB}}\right)   \label{bobpnb2}
\end{eqnarray}%
to solve the Schr$\ddot{\rm o}$dinger equation of the light atom.
Here, $r_{iB}$ is the distance between the heavy atom $A_{i}$ and the light atom $B$.
That is the approach we used in our main text.

\end{subappendices}

%%%%%%%%%%%%%%%%%%%%%%%%%%%%%%%%%%%%%%
%%%%%%%%%%%%%%%%%%%%%%%%%%%%%%%%%%%%%%
%%%%%%%%%%%%%%%%%%%%%%%%

%%%%%%%%%%%%%%%%%%%%%%%%%%%%%%%%%%%%%%%%%%%
%%%%%%%%%%%%%%%%%%%%%%%%%%%%%%%%%%%%%%%%%%%
%%%%%%%%%%%%%%%%%%%%%%%%%%%%%%%%%%%%%%%%%%%
%%%%%%%%%%%%%%%%%%%%%%%%%%%%%%%%%%%%%%%%%%%
%%%%%%%%%%%%%%%%%%%%%%%%%%%%

\end{document}